\documentstyle[11pt,newpasp,twoside,epsf]{article}
\def\etal{{et\thinspace al.}\ }
\def\logg{$\log g$\thinspace}

\newcommand{\Teff}{\hbox{$T_{\rm eff}$}}

\newcommand{\ion}[2]{\mbox{#1\,{\small #2}}}

\begin{document}
\title{Temperature Scale and Iron Abundances of Very Hot Central Stars of Planetary Nebulae}
 \author{Klaus Werner, Jochen L. Deetjen, Stefan Dreizler, Thomas Rauch}
\affil{Institut f\"ur Astronomie und Astrophysik, Univ.\ T\"ubingen, Germany}
\author{Jeff W. Kruk}
\affil{Department of Physics and Astronomy, JHU, Baltimore MD, U.S.A.}

\begin{abstract}
The determination of effective temperatures of very hot central stars
(\Teff$>70\,000$\,K) by model atmosphere analyses of optical H and He line
profiles is afflicted with considerable uncertainty, primarily due to the lack
of neutral helium lines. Ionization balances of metals, accessible only with UV
lines, allow more precise temperature estimates. The potential of iron lines is
pointed out. At the same time iron and other metal abundances, hardly
investigated until today, may be derived from UV spectra. We describe recent
HST spectroscopy performed for this purpose.

A search for iron lines in FUV spectra of the hottest H-deficient central stars
(PG1159-type, \Teff$>$100\,000\,K) taken with FUSE was unsuccessful. The
derived deficiency is interpreted in terms of iron depletion due to n-capture
nucleosynthesis in intershell matter, which is now exposed at the stellar
surface as a consequence of a late He shell flash.
\end{abstract}

\section{Introduction}

In this review we choose to concentrate on very hot central stars of planetary
nebulae (CSPN), whose analysis faces several difficulties. These objects
generally have weak or very weak winds and can be analyzed with windless model
atmospheres. Two other reviews in these proceedings concentrate on analyses of
CSPN with strong winds (Hamann and Pauldrach for H-deficient and H-rich stars,
respectively).

CSPN are spectroscopically assigned to one of two families: H-rich or
H-deficient. The origin of H-deficiency is now regarded as the consequence of a
(very) late helium shell flash occurring during post-AGB evolution (Herwig
\etal 1999) and we observe a complete H-deficient post-AGB sequence. On the
other hand, the apparent gap in the separate H-rich sequence has been closed by
Napiwotzki (1999) who discovered numerous  very hot (white dwarf) CSPN.

Several spectral sub-classes for the two sequences have been established by
M\'endez (1991). Most of his terms are still used in present work, some new have
been added by others. To start our discussion we summarize these classification
criteria.

\section{H-rich and H-deficient CSPN}

The H-rich CSPN can be classified by the following spectral characteristics.

hgO(H): denotes high-gravity stars with very broad Balmer absorptions. These
are either hot white dwarfs or non-post-AGB stars.

O(H): stars with spectra very similar to massive, young O stars, with
\ion{He}{II}~4686\AA\ in absorption. Spectral subtypes can be defined from the
relative strengths of \ion{He}{I} and \ion{He}{II} absorption, but excellent
spectra are needed, because \ion{He}{I} absorptions are often masked by nebular
emission. The hottest CSPN do not show \ion{He}{I} absorptions at all.

Of(H): the stellar \ion{He}{II}~4686\AA\ line is a narrow emission
(FWHM\,$>$\,4\AA, sometimes with P~Cyg profile). \ion{He}{II}~4200,\,4541\AA\
are in absorption, and blueshifted by less than 50\,km/s relative to the PN
velocity. H$\gamma$ is in absorption.

Of-WR(H): \ion{He}{II}~4686\AA\ is a strong and broad emission, H$\gamma$ shows
a more or less developed P~Cyg profile. \ion{He}{II}~4541\AA\ is not in
emission but, if visible in absorption, strongly blueshifted (about 100\,km/s).

Spectral analysis of these stars has been pioneered by Kudritzki, M\'endez and
co-workers (e.g.\ M\'endez \etal 1988), by determining basic photospheric
parameters (\Teff, \logg, He abundance). Subsequently, more analyses in this
style have been published, too numerous to refer to here. Napiwotzki (1999)
published new basic results on many CSPN including a very useful compilation of
CSPN analyses with an extensive reference list. Mass-loss rates of O- and
Of-type CSPN have been determined, too (e.g.\ Kudritzki \etal 1997). Metal
abundance analyses of H-rich CSPN are rather scarce and mostly qualitative
descriptions of spectral signatures were performed. We discuss this in more
detail below. Recent and current research on H-rich CSPN  concentrates on wind
modeling (Pauldrach, these proceedings) for the O and Of types, and on problems
with \Teff\ determination for the hottest O and hgO types (Napiwotzki 1999, and
the present paper), and on metal abundance determinations in the latter (see
below).

The H-deficient CSPN are classified as follows.

O(He), O(C): Stars showing predominantly an absorption line spectrum, dominated
by He or C, respectively. O(C) stars are now more commonly called PG1159 stars
(after PG1159-035) although NGC\,246 is the prototype of this class (Heap
1975). The PG1159 classification has been further refined, using line-shapes of
the dominant species as characteristic criteria (Werner 1992).

Of(C), Of-WR(C): Defined in analogy to the H-rich counterparts (without meeting
criteria for presence of H). These stars are also termed [WC]--PG1159
transition objects because of their mixed emission/absorption line spectra,
bearing characteristics of both spectral classes. Examples are Abell~30 and
Abell~78.

[WC]: Defined like in the case of massive Wolf-Rayet stars, with necessary
extensions to earlier and later subtypes.

We remark that all of these stars are sometimes called ``\ion{O}{VI}'' CSPN,
because of the prominence of an \ion{O}{VI} doublet in the optical UV. The
so-called ``weak emission line stars'' (WELS) are sometimes equated to the
[WC]--PG1159 objects (Parthasarathy \etal 1998) but better spectra and
quantitative analyses  may reveal that many of them are either [WC] or PG1159
stars, or even H-rich CSPN with only slightly enriched metal abundances (like
discussed below).

Some PG1159 and [WC] stars do show photospheric hydrogen, which is difficult to
detect because of high temperatures, blending \ion{He}{II} photospheric or H
nebular lines. H abundances in PG1159 stars can be as high as 35\% (by mass),
placing the respective objects somewhat between the H-rich and H-deficient
CSPN, hence, they are termed hybrid-PG1159 stars. However, from an evolutionary
standpoint they are to be assigned to the H-deficient sequence, because of
their related origin (so-called ``AGB Final Thermal Pulse'' scenario, Herwig
2001). The O(He) stars (e.g.\ K1-27) are very rare (two CSPN and two similar
objects without PN) and their status and relation to other post-AGB stars is
not really understood (Rauch \etal 1998).

During the last decade, considerable effort has been put into analyses of
H-deficient CSPN, see the review on [WC] stars (Hamann, these proceedings) and
a recent summary on PG1159 stars (Werner 2001). It is now well established that
both subclasses represent different stages of an evolutionary sequence.

Considerable progress, which can only be shortly mentioned here, has been
achieved with NLTE atmospheric modeling since the last PN symposium. For models
with winds we refer again to the reviews by Pauldrach and Hamann in these
proceedings. Windless models are now available over a wide parameter range,
including opacities of light metals and iron group elements (see contribution
by Rauch, these proceedings). Results reported here were obtained with such
models for both H-deficient and H-rich objects. They are calculated with an
Accelerated Lambda Iteration code (Werner \& Dreizler 1999).

\section{Very hot H-rich CSPN}

On the PN symposium in Innsbruck, Liebert (1993) pointed out that ``{\it
ironically, the temperature determinations for the H-poor sequence with their
complicated and mixed atmospheric abundances of He and CNO elements may
presently be more accurate than those of the H-rich sequence.}'' This was
certainly true for hot CSPN (say, \Teff$>$70\,000\,K), particularly because the
``Balmer line problem'' came up at that time (Napiwotzki 1992). The lack of
neutral helium lines in these stars prevents a reliable \Teff\ determination
based on the \ion{He}{I}/\ion{He}{II} ionization balance, hence, the Balmer line
spectrum alone was used for that purpose. But it turned out that no consistent
fit to all lines of the Balmer series could be achieved: The higher the series
member, the higher the required fit temperature. Although we have learned to
circumvent this problem (the higher series member give the ``correct'' \Teff)
and partly understand its reason (neglect of metal line blanketing in the NLTE
models, Werner 1996), considerable problems remain for the hottest CSPN. In
contrast to cooler CSPN, line blanketed models do not resolve the Balmer line
problem for these stars. The reason is still not understood, hence, other means
must be used for determining their temperatures. These ``other means'' are the
ionization balances of metals.

\begin{figure}[ht]
\epsfxsize=7.5cm
\epsffile{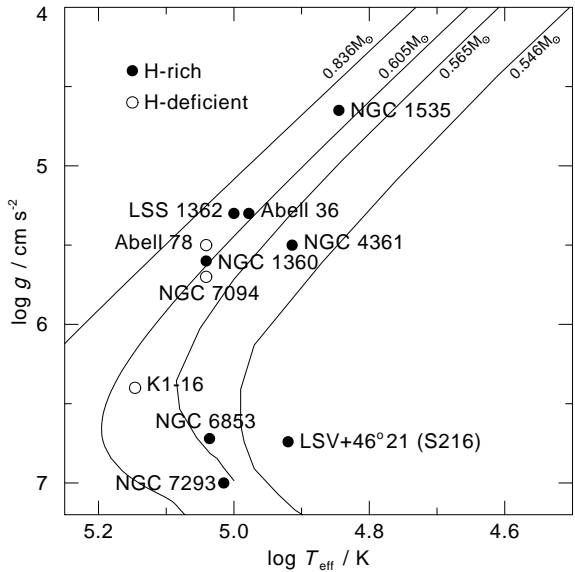}
\caption{
Location of the very hot CSPN in the $g$--\Teff\ plane subject to HST and FUSE
spectroscopy as described in the text. Evolutionary tracks for H-burning
post-AGB stars with different remnant masses are from Sch\"onberner (1983) and
Bl\"ocker \& Sch\"onberner (1990).
}\label{hrd}
\end{figure}

\begin{figure}[ht]
\epsfxsize=10cm
\epsffile{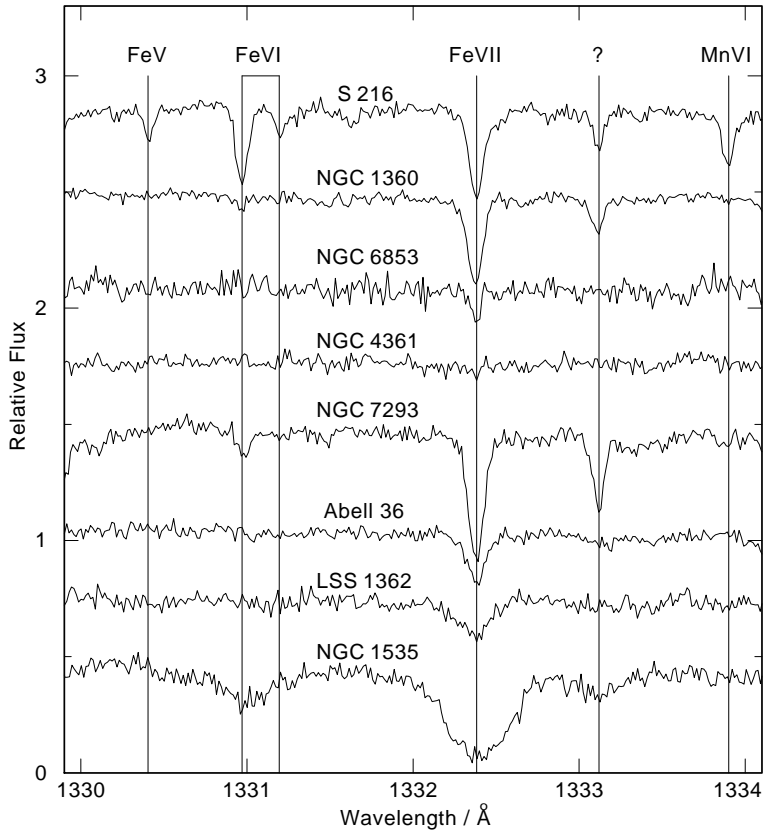}
\caption{
Detail from HST spectra of very hot CSPN, displaying lines from
\ion{Fe}{V-VII}. Their relative strength is an indicator for \Teff. Also seen
are a \ion{Mn}{VI} line in S\,216 and an unidentified absorption that possibly
stems from an as yet unknown \ion{Fe}{VII} line.
}\label{hst_spectra}
\end{figure}

Metals in very hot CSPN are highly ionized and the strongest lines of these
ions are found in the UV region. UV high resolution spectra usually display
lines from many metals so that a number of ionization balances can be exploited
simultaneously in order to fix \Teff. This is highly desirable, because in
principle other photospheric parameters which also affect ionization balances
(\logg\ and element abundances) must be derived from these features at the same
time. Iron promised to be exceptionally useful, because iron lines in IUE
spectra of hot CSPN and related subdwarfs were reported earlier (Sch\"onberner
\& Drilling 1985). In order to proof the viability of such an analysis we
proposed HST UV spectroscopy of eight H-rich CSPN. Their location in the
$g$--\Teff\ diagram (mainly derived from optical analyses and, hence, possibly
uncertain) is displayed in Fig.\,\ref{hrd}. Another immediate motivation for
these observations came from an analysis of archival IUE spectra, which
indicated a possible general iron deficiency in many hot H-rich CSPN (Deetjen
\etal 1999). Furthermore, M\'endez (1991) emphasized that the optical spectra
of many hot H-rich CSPN indicate a variety of carbon abundances which promise
insight into previous (post-) AGB evolutionary processes. Our HST observations
were performed in 2001/2002 with the STIS spectrograph and cover the wavelength
range 1150-1730\AA\ with a resolution of better than 0.1\AA\ and
S/N$\approx$50. The data are currently analyzed. Only NGC\,1535 exhibits wind
signatures in the strongest CNO lines. Sample spectra shown in
Fig.\,\ref{hst_spectra} give an impression of the data quality. We discuss some
qualitative results on individual objects in some detail.

\begin{figure}[ht]
\epsfxsize=\textwidth
\epsffile{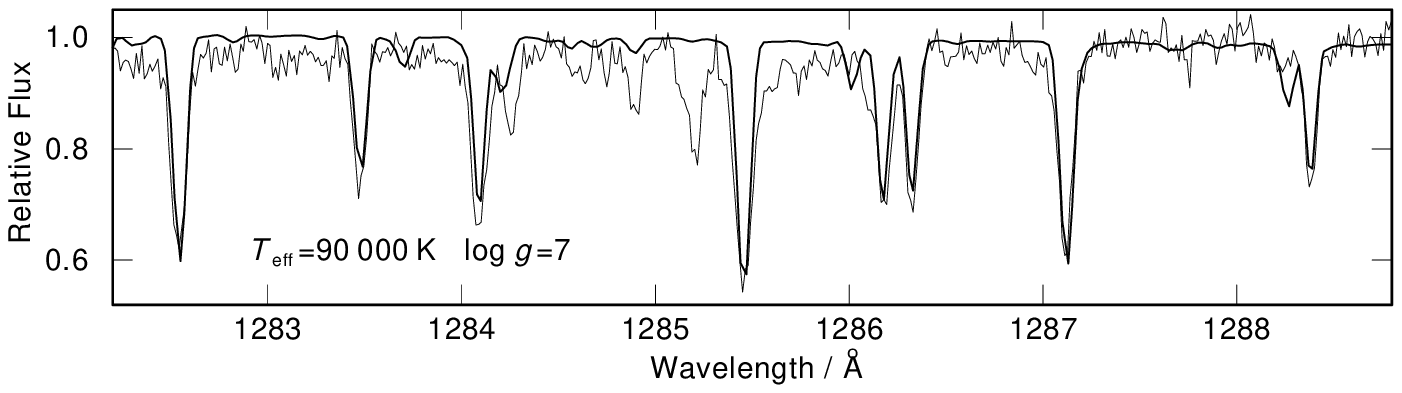}
\caption{
Preliminary fit to the \ion{Fe}{VI} line spectrum of S\,216 with a solar iron
abundance model. The unmatched absorptions at 1284.14\AA\, 1284.79\AA,  and
1285.10\AA\ are two \ion{Ni}{VI} lines and a \ion{Mn}{VI} line, respectively,
which are not included in the model.
}\label{s216}
\end{figure}

NGC\,4361 was highlighted by M\'endez (1991), because it shows very strong
\ion{C}{IV} lines in the optical, as compared to NGC\,1360 which has otherwise
similar \Teff\ and \logg\ thus, NGC\,4361 is apparently strongly overabundant
in C. In the meantime, we have presented evidence from extended optical and EUV
observations (Hoare \etal 1996) that NGC\,1360 is about 20\,000\,K hotter than
previously thought, which possibly explains the relative weakness of
\ion{C}{IV} in this star. But still, the strong \ion{C}{IV} lines in NGC\,4361
are even more remarkable since Torres-Peimbert \etal (1990) found that it must
be a metal-poor halo object. Our HST spectrum confirms both, the halo nature of
NGC\,4361 because of the almost complete absence of iron lines, and the high C
abundance: NGC\,4361 reveals the strongest \ion{C}{IV} lines in our sample of
stars. The probably high C overabundance is reminiscent of the same situation
in K\,648, the central star of the PN Ps\,1 in the globular cluster M15 (Rauch
\etal 2002). NGC\,1360 on the other hand shows a rich spectrum of
\ion{Fe}{VI}/\ion{Fe}{VII} lines. They are weaker than expected, which either
indicates an even higher \Teff\ or a subsolar Fe abundance.

LSS\,1362 and Abell\,36 show only \ion{Fe}{VII} but no \ion{Fe}{VI} lines. This
might indicate that these stars are hotter than NGC\,1360, as opposed to their
currently known position in Fig.\,\ref{hrd}. This is supported by the fact that
all three stars display \ion{O}{V} and \ion{O}{VI} lines while only NGC\,1360
displays \ion{O}{IV}, too.

NGC\,6853 and NGC\,7293 both have high gravities and do show iron lines.
NGC\,7293 has \ion{Fe}{VI} and \ion{Fe}{VII} lines, NGC\,6853 only
\ion{Fe}{VII}. The higher \Teff\ and the lower gravity of NGC\,6853 explains
this qualitative difference. The helium abundance in NGC\,6853 is solar, while
it is subsolar in NGC\,7293, indicating that the photospheric abundances are
influenced by gravitational settling in the latter. Detailed analyses are
needed to see if metal abundances comprise the signature of diffusion and
radiative levitation. The white dwarf central star of S\,216 has a spectrum
which is very rich in iron (\ion{Fe}{V-VII}) and nickel (\ion{Ni}{V-VI}) lines
(see Fig.\,\ref{s216}), the strongest of them were detected earlier in IUE
spectra (Tweedy \& Napiwotzki 1992). In our HST spectra we also find lines of
other iron group elements: chromium, cobalt, and manganese (ionization stages V
and VI). Their line strengths as well of Ni clearly points at an overabundance
with respect to Fe, which is a result of radiative levitation. Many absorptions
in the UV remain unidentified, most of them are probably unknown iron lines, as
can be concluded by comparison with known line features in the different stars.

From the FWHM of the iron lines in NGC\,1360 and NGC\,7293 we find that the
projected equatorial rotation velocity or any turbulent atmospheric motion does
not exceed about 20\,km/s. This is in contrast to M\'endez \etal (1988) who
deduced 100\,km/s and 50\,km/s, respectively, from weak metallic absorption
lines in their optical spectra, and who attributed this to some kind of
``macroturbulence''. Similar iron line widths are found in  S\,216 and
NGC\,6853. For Abell\,36 and LSS\,1362 we find larger values, namely 30\,km/s
and 50\,km/s.

NGC\,1535 is the lowest gravity CSPN in our sample. It has a most remarkable UV
spectrum. The strongest CNO lines have P~Cygni profiles which has been know
long before (Heap 1983). What is new, however, is the detection of prominently
strong and broad \ion{Fe}{VI} and \ion{Fe}{VII} lines
(Fig.\,\ref{hst_spectra}). We think that this is no indication for an iron
overabundance. Rather, the broad lines indicate some macroturbulent motions
with a speed of the order of 100\,km/s. Clearly, this object requires an
analysis with atmosphere models including a wind.

\section{Very hot hydrogen-deficient CSPN}

The first attempt to determine the iron abundance in H-deficient CSPN arrived
at a surprising result. Since PG1159 stars belong to the disk population and
since diffusion effects can be excluded due to ongoing mass-loss, one expects a
solar iron abundance in these stars. However, Miksa \etal (2001) report an
underabundance of at least one dex as derived from the absence of \ion{Fe}{VII}
lines in FUSE spectra of the extremely hot CSPN K1-16. Subsequent analyses of
FUSE spectra from other PG1159 stars confirmed that such an Fe underabundance
is possibly the rule rather than the exception among these objects: Miksa \etal
(2002) and Werner \etal (these proceedings) failed to detect \ion{Fe}{VII}
lines in NGC\,7094 and Abell~78, respectively.

As already mentioned, the high C and O abundances in H-deficient CSPN results
from envelope mixing caused by a late He-shell flash, hence, this event also
dredges up matter where s-process elements were built up by n-capture on
$^{56}$Fe seeds during the AGB phase. In principle, this scenario can be tested
by analyzing the resulting Fe/Ni abundance ratio, because it is significantly
changed in favor of Ni by the conversion of $^{56}$Fe into $^{60}$Ni. The Fe
depletion by n-captures typically amounts to a factor of 10 (Busso \etal 1999).
In order to roughly estimate the Ni/Fe ratio one can assume nuclear statistical
equilibrium. The two most abundant Ni isotopes are $^{60}$Ni (26\%) and
$^{58}$Ni (68\%). During s-process $^{58}$Ni is destroyed (and not
synthesized), by conversion into $^{60}$Ni. $^{60}$Ni is converted to heavier
elements 4 times faster than it is produced from $^{56}$Fe. Consequently, a
ratio Fe/Ni$\approx$4 results, which is a factor of five below the solar value.
This could be detected by high resolution HST and FUSE UV spectroscopy.
Interestingly, Asplund \etal (1999) have indeed found that in Sakurai's object,
which is thought to undergo a late He-shell flash, Fe is reduced to 0.1 solar
and Fe/Ni$\approx$3. This and other s-process signatures might also be
exhibited by Wolf-Rayet central stars and PG1159 stars. More quantitative
results from nucleosynthesis calculations in appropriate stellar models have
been presented at this conference by Herwig \etal\ and inclusion of nuclear
networks in evolutionary model sequences will become available in the near
future.

Other results presented at this conference confirm that Fe deficiency among
H-deficient post-AGB stars is not restricted to PG1159 stars, as can be
expected from evolutionary considerations. Three Wolf-Rayet central stars are
iron deficient, too. Gr\"afener \etal report a low Fe abundance in SMP\,61, an
early type [WC5] central star in the LMC. Its abundance is at least 0.7\,dex
below the LMC metallicity. Crowther \etal find evidence for an iron
underabundance of 0.3--0.7\,dex in the Galactic [WC] stars NGC~40 ([WC8]) and
BD+30\,3639 ([WC9]).

\section{Conclusion}

We have discussed the potential of high-resolution UV spectroscopy for the
determination of photospheric parameters of the hottest CSPN. Ionization
equilibria of metals, particularly of iron, are the only means for precise
temperature determinations. NLTE model atmospheres have been developed to fully
exploit this technique. HST and FUSE are essential instruments and have been
successfully used to obtain F(UV) spectra of the necessary high quality.

Metal abundance determinations from UV data are necessary to understand mixing
processes in the interior of the precursor AGB stars. In the case of
hydrogen-deficient CSPN the surface abundances reflect the intershell
abundances of the former AGB star with some superimposed modification due to
the mixing and nucleosynthesis processes as a consequence of a late thermal
pulse. This can also explain their Fe deficiency, as due to n-capture
nucleosynthesis on Fe seeds during the AGB evolution and/or the neutron burst
which occurs during rapid burning of protons initiated by the late He shell
flash.

\begin{acknowledgements}
HST data analysis in T\"ubingen is supported by the DLR under grant
50\,OR\,9705\,5.
\end{acknowledgements}

\end{document}